\documentclass[preprint,12pt]{elsarticle}

\newcommand{\isdef}{\stackrel{\triangle}{=}}

\usepackage{amssymb}

\usepackage{epsfig,subfigure,pstcol}

\journal{ASME Journal of Applied Mechanics}

\begin{document}

\begin{frontmatter}



\title{Constitutive-law Modeling of Microfilaments from their Discrete-Structure Simulations - A Method based on an Inverse Approach Applied to a Static Rod Model}


\author{Adam R. Hinkle}

\address{Theoretical and Applied Mechanics\\
	Cornell University\\
	Ithaca, NY 14853-1503\\
    Email: ah353@cornell.edu}
    
\author{Sachin Goyal}

\address{Department of Physics\\
	Emory University\\
	Atlanta, GA 30322-2430\\
    Email: sgoyal@umich.edu}
    
\author{Harish J. Palanthandalam-Madapusi}

\address{Mechanical and Aerospace Engineering\\
	Syracuse University\\
	Syracuse, NY 13244-1240\\
    Email: hjpalant@syr.edu}

\begin{abstract}
Twisting and bending deformations are crucial to the biological functions of several microfilaments such as DNA molecules. Although continuum-rod models have emerged as efficient tools to describe the nonlinear dynamics of these deformations, a major roadblock in the continuum-mechanics-based description of microfilaments is the accurate modeling of the constitutive law, which follows from its atomistic-level structure and interactions. Since first-principle derivation of the constitutive law from atomistic-level structure and interactions is often impractical and so are direct experimental measurements due to the small length-scales, a natural alternative is to estimate the constitutive law from discrete-structure simulations such as molecular-dynamics (MD) simulations. In this paper, we describe a method based on a recently proposed two-step inverse approach \citep{palanth:10a} for estimating the constitutive law using a static rod model and deformed configuration data generated from discrete-structure simulations. Furthermore, we illustrate the method on a filament with an artificial discrete-structure. We simulate its deformation in response to a prescribed loading using a multi-body dynamics (MBD) solver, Hyperworks MotionView. Using position data generated from the MBD solver, we first estimate the curvature of the filament, and subsequently use it in the described method to estimate the effective constitutive-law relationship between the restoring moment and curvature. Finally, we also illustrate how the estimated constitutive law can be validated under independent loading conditions.

\end{abstract}

\begin{keyword}
Microstructure, constitutive behavior, rod theory,  biological material, inverse methods.


\end{keyword}

\end{frontmatter}


\section{INTRODUCTION}
\label{intro}

Microfilaments represent a wide range of natural or synthetic microscopic structures that are characterized by a large length-to-width aspect ratio and exhibit large bending and twisting deformations. Examples of microfilaments include DNA molecules, microtubules, flagella, fimbriae and setae. The functions of these biological microfilaments are directly affected by their large nonlinear deformations and therefore it is important to understand and model the dynamics of these deformations. Although the ideas and methods discussed in this paper are applicable to almost all microfilaments, our motivation stems from the fact that the biological functions of DNA molecules such as gene expression, transcription, and replication are significantly influenced by its long length-scale structural deformations such as looping \citep{schleif:92a,semsey:05a}. Recent successes of continuum-rod models in describing biologically-relevant deformations of DNA molecules \citep{goyal:07a,balaeff:06a} suggest that a continuum-rod model is an efficient tool to describe deformations of microfilaments.

A continuum-rod model in general consists of dynamic equilibrium equations and compatibility equations, which needs to be solved respecting a constitutive law. Although dynamic equilibrium equations and compatibility equations, which we henceforth refer to as the {\it rod model}, remain the same for all microfilaments, the key distinguishing factor is the constitutive law. For each microfilament, the constitutive law, which describes the relationship between the deformation of the microfilament and the restoring forces and moments, follows from its atomistic structure and bond stiffnesses.

The rod model equations, that is, the equilibrium and compatibility equations, reviewed in Section \ref{rod model}, can be recognized as four vector partial differential equations in space $s$ and time $t$ \citep{goyal:05b}. In this paper, we assume the microfilament to be inextensible and unshearable. Under these assumptions, the four rod model equations involve five vector unknowns, namely, internal force $\vec{f}(s,t)$, internal moment $\vec{q}(s,t)$, curvature vector $\vec{\kappa}(s,t)$, velocity $\vec{v}(s,t)$, and angular velocity $\vec{\omega}(s,t)$. The constitutive law serves as the fifth vector algebraic equation that captures the restoring effect of the internal force  $\vec{f}$ and internal moment $\vec{q}$ on the curvature $\vec{\kappa}$. A general form of an elastic constitutive law applicable to an inextensible and unshearable rod model (discussed in Section \ref{constitutive law}) is
\begin{equation}
\vec{\psi} \left( \vec{\kappa}, \vec{q}, \vec{f}, s \right) = 0,
\label{const}
\end{equation}
where $\vec{\psi}$ is a vector function of its arguments and follow from atomistic-level interactions. While the rod model equations can be derived analytically from first principles, first-principle derivation of the constitutive law from atomistic-level interactions is often impractical.

Alternatively, if all the arguments of $\vec{\psi}$, namely, internal force, internal moment, and curvature vectors at each cross section, are experimentally measured, then well-established function-approximation tools can be used to estimate the constitutive law. Although such an approach will help circumvent approximations due to modeling assumptions, it may introduce experimental artifacts. Nonetheless, while such measurements may be conceivable with acceptable accuracy for some laboratory-scale rods, the technology does not yet exist for sub-micron filaments. One of the many challenges at small length scales is that thermal fluctuations may dominate and thus corrupt any potential measurements. Some experiments exploit thermal fluctuations to measure persistence lengths and thus estimate average bending and torsional stiffnesses, which are dominant parameters in a linearized, homogeneous approximation of equation (\ref{const}); for example, refer to \citep{hagerman:88a,strick:96a} for such experiments on DNA molecules. However, such simplistic constitutive laws are often inadequate in modeling the deformation of microfilaments \citep{cloutier:04a,wiggins:05a,goyal:08b,smith:08a}.

A more practical alternative than using experimental measurements is to use data generated from discrete-structure simulations such as molecular dynamics (MD) simulations, which account for atomistic-level structure and interactions, and are accurate and computationally feasible for short length scales \citep{schlick:02a}.  This alternative is already gaining attention for modeling constitutive laws of several biological materials such as microtubules \citep{arslan:10a} and DNA \citep{lankas:09a}, although the attempts have been limited to estimation of only a few parameters, wherein the form of the constitutive law is presumed {\it a priori}, and is often presumed linear. Recently, a two-step inverse approach has been proposed \citep{palanth:10a} that can estimate directly the vector function $\vec{\psi}$ from the discrete-structure simulations. This approach exploits state-space form of a static rod model, and can leverage several different scenarios of input and state measurements for the constitutive-law estimation.

In this paper, we describe a method based on the two-step inverse approach \citep{palanth:10a} for a scenario, when the equilibrium positions of individual atoms computed from discrete-structure simulations under judiciously chosen loading conditions can provide the necessary information for estimating the effective constitutive law. In this method, which is described in Section \ref{estimation}, we first estimate the curvature $\vec{\kappa}$ from the position data of the deformed configuration generated from discrete-structure simulations. This pre-processing involves numerical differentiations and pose associated challenges \citep{ramm:04a}, which often necessitate smoothing of numerical data \citep[Appendix B]{goyal:07a}. Next, following the step one of \citep{palanth:10a}, we describe how the estimated curvature $\vec{\kappa}$ can be used as a known quantity in the static rod model equations to solve for the internal restoring force  $\vec{f}$ and internal restoring moment $\vec{q}$ in static equilibrium of the deformed configuration. Finally, following the step two of \citep{palanth:10a},, a least-squares fit can be used to estimate the functional relationship $\vec{\psi}$ between the estimated curvature $\vec{\kappa}$, the estimated force $\vec{f}$, and the estimated moment $\vec{q}$.

In Section \ref{results}, we present results illustrating the applicability of the method on a filament with an artificial discrete-structure, which we simulate using a multi-body dynamics (MBD) solver. The artificial discrete-structure is so chosen that the expected constitutive law is homogeneous and has no dependence on internal force $\vec{f}$. Furthermore, the artificial discrete-structure has a decoupled behavior in two-axes bending and torsion. So, it suffices to present the viability of the proposed method for planar bending about one principal bending axis. The simplicity of the artificial discrete-structure also allows us to offer a first-principle explanation of the estimated constitutive law as further corroboration. We, however, emphasize that the method described in Section \ref{estimation} is versatile, but illustrating its application to a variety of general cases is beyond the scope of this paper.

To put our method in perspective, we begin by first reviewing a general-purpose 3-D formulation of the rod model \citep{goyal:05b} in Section~\ref{rod model}, which simulates the dynamics of large, nonlinear, deformations including intertwining \citep{goyal:08c}. In Section \ref{constitutive law}, we introduce a general, conceivable form of the constitutive law for an inextensible and unshearable microfilament. In Section~\ref{estimation}, we describe the method for the estimation of the constitutive law. The method exploits the static formulation of the rod model reviewed in Section~\ref{rod model}. In Section \ref{results}, we present and discuss the results illustrating the estimation method on an artificial discrete-structure. We finally close by summarizing the conclusions in Section \ref{conclusions}.

\section{The Rod Model}
\label{rod model}

\begin{figure}[h!]
 \centering
 \includegraphics{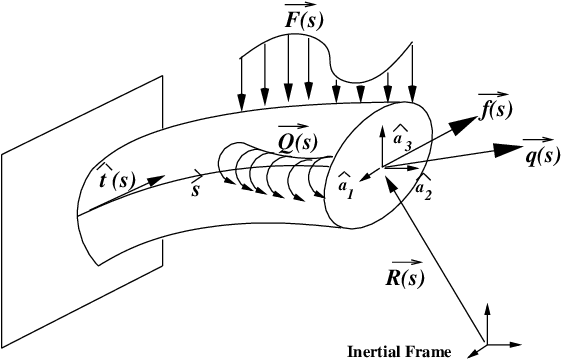} 
  \caption{Rod model of a microfilament in dynamic equiilibrium illustrating quantities of interest.}
 \label{fig:dnarod}
 \end{figure}

In this section, we review the mathematical formulation of the rod model \citep{goyal:05b} that involves dynamic equilibrium equations and compatibility conditions. In the rod model, the dynamics of the microfilament deformation follows from the rigid-body motion of its individual cross-sections. To track the rigid body motion of each cross-section, we fix at its mass center a reference frame $\hat{a}_{i}(s,t)$, where the subscript $i = 1,2,3$, the independent variable $s$ is the unstretched centerline coordinate and the independent variable $t$ is time. As shown in  Figure~\ref{fig:dnarod}, we denote the position of the cross-section-fixed reference frame by $\vec{R}(s,t)$. The gradient of $\vec{R}(s,t)$ along $s$, which we denote by $\vec{r}(s,t)$, points along the centerline tangent $\hat{t}(s,t)$. Rod deformation in general involves two-axes-bending and torsion captured by a curvature vector $\vec{\kappa}(s,t)$ as well as shear along the two flexure axes and extension (or compression) captured by $\vec{r}(s,t)$. In the stress-free (undeformed) state $\vec{r}(s,t) = \hat{t}(s,t)$ and we denote the stress-free curvature by $\vec{\kappa_{0}}(s)$. A change in the magnitude of $\vec{r}(s,t)$ captures extension (or compression), while change in its orientation with respect to $\hat{a}_{i}(s,t)$ captures shear. The stress distribution across the cross-section results in a net internal (tensile and shear) force $\vec{f}(s,t)$ and (bending and torsional) moment $\vec{q}(s,t)$. The rigid body motion of the cross-section is described by its translational velocity $\vec{v}(s,t)$ and its angular velocity $\vec{\omega}(s,t)$.

The governing equations for the rod dynamics are
\begin{eqnarray}
\frac{\partial \vec{f}}{\partial s} + \vec{\kappa} \times \vec{f} &=& m\frac{\partial \vec{v}}{\partial t} + m \vec{\omega} \times \vec{v} - \vec{F}, \label{linear_momentum} \\
\frac{\partial \vec{q}}{\partial s} + \vec{\kappa} \times \vec{q} &=& {\bf{I}}\cdot\frac{\partial \vec{\omega}}{\partial t} + \vec{\omega} \times {\bf{I}}\cdot\vec{\omega} + \vec{f} \times \vec{r} - \vec{Q}, \label{angular_momentum}\\
\frac{\partial \vec{v}}{\partial s} + \vec{\kappa} \times \vec{v} &=& \frac{\partial \vec{r}}{\partial t} + \vec{\omega} \times \vec{r}, \label{position_continuity} \\
\frac{\partial \vec{\omega}}{\partial s} + \vec{\kappa} \times \vec{\omega} &=& \frac{\partial \vec{\kappa}}{\partial t}, \label{orient_continuity}
\end{eqnarray}
where $m(s)$ is the mass of the rod per unit length, the tensor ${\bf{I}}(s)$ is the mass moment of inertia per unit length,  $\vec{F}(s,t)$ is the distributed external force per unit length, and  $\vec{Q}(s,t)$ is the distributed external moment per unit length. The derivatives are all relative to the rotating, cross-section-fixed reference frame $\hat{a}_{i}(s,t)$. The dynamic equilibrium equations (\ref{linear_momentum}) and (\ref{angular_momentum}) can be derived by applying Newton's second law to an infinitesimal rod element. The compatibility equations (\ref{position_continuity}) and (\ref{orient_continuity}) follow from the space-time continuity of the cross-section position $\vec{R}(s,t)$ and the cross-section orientation $\hat{a}_{i}(s,t)$ respectively \citep[Appendix 2]{goyal:05b}. A physical interpretation of these compatibility equations is that the gradient of the translational and angular velocities captures the time-rate of change of the rod's deformation.

The rod model described above has been successfully exploited to offer promising insights on sequence-dependent DNA looping \citep{goyal:07a}, kinking \citep{goyal:08b}, intertwining \citep{goyal:08c} and buckling of microfibril arrays \citep{naderman:10a}, albeit assuming simplistic constitutive laws. Next, we discuss the constitutive law.

\section{Constitutive Law for The Rod Model}
\label{constitutive law}

In this paper, we consider the rod model for slender microfilaments under low-tension applications that dominantly involve bending and twisting deformations. We assume that the microfilaments are inextensible and unshearable. This assumption leads to the simplification $\vec{r}(s,t) = \hat{t}(s,t)$, which is then a constant unit vector as seen from the cross-section-fixed reference frame $\hat{a}_{i}(s,t)$ \citep{goyal:05c}. Then, the rod model equations (\ref{linear_momentum}) - (\ref{orient_continuity}) constitute a set of four vector partial differential equations in five vector unknowns, namely translational velocity $\vec{v}(s,t)$, angular velocity $\vec{\omega}(s,t)$, internal force $\vec{f}(s,t)$, internal moment $\vec{q}(s,t)$, and curvature vector $\vec{\kappa}(s,t)$. The nonlinear rod model equations (\ref{linear_momentum}) - (\ref{orient_continuity}) can be solved numerically along with a fifth vector equation, the constitutive law. A general form of an elastic constitutive law applicable to an inextensible and unshearable rod model is the algebraic constraint
\begin{equation}
\vec{\psi} \left( \vec{\kappa}, \vec{q}, \vec{f}, s \right) = 0.
\label{eq:const}
\end{equation}

The constitutive law (\ref{eq:const}) captures the restoring effects of internal force $\vec{f}(s,t)$ and internal moment $\vec{q}(s,t)$ on curvature vector $\vec{\kappa}(s,t)$ due to the stress-strain relationship integrated over the rod's cross section.  It is worthwhile to note that ignoring shear and extension does not ignore their kinematic coupling with twisting and bending. This coupling is captured by internal force $\vec{f}(s,t)$ in the constitutive law (\ref{eq:const}). An example of a constitutive law with twist-extension coupling in inextensible rod model is presented in \citep[Section 6]{goyal:05b}.

The elastic behavior of microfilaments and hence the form of their constitutive law follow from their atomistic-level interactions. In fact, there is no clear consensus on its functional form for DNA molecules and how it maps from the base-pair sequence. Although several experiments have provided estimates of average torsional and bending stiffness by measuring persistence lengths \citep{hagerman:88a,strick:96a}, recent experimental observations \citep{cloutier:04a} and model-fitting \citep{wiggins:05a,smith:08a} suggest that the constitutive law may be nonlinear, wherein the kink-ability of the molecule indicate non-convex stored energy functions \citep{fosdick:81a}.

Next, we present a general-purpose two-step computational method that can exploit the rod model equations along with data generated from discrete-structure simulations to estimate the vector function $\vec{\psi}$.

\section{Two-Step Estimation of Constitutive Law from Atomistic Structure Simulations}
\label{estimation}

To estimate the effective constitutive law resulting from an atomistic pattern, we consider a cantilever microfilament of length $L$ with several repeats of the atomistic pattern, and consider its deformation in {\it static equilibrium}. The cantilever is clamped at one end (at $s=L$) and body-fixed forces and moments are applied to the other end (at $s=0$).  This loading arrangement allows us to use a static formulation of the rod model, along with the data generated from discrete-structure simulations, in a two-step inverse approach \citep[Table 1, Scenario 1]{palanth:10a} to estimate the effective constitutive law. For employing the two-step method, we write the static rod model equations in state-space form, which provides a general-purpose framework for a variety of estimation scenarios as elaborated in \citep[Table 1]{palanth:10a}.

\subsection{Static Rod Model}
\label{static rod}

In static equilibrium, the rod model reduces to a time-independent formulation, that is, translational velocity $\vec{v}$, angular velocity $\vec{\omega}$ and all partial derivatives with respect to $t$ in equations (\ref{linear_momentum}) - (\ref{orient_continuity}) vanish. As a result, the compatibility equations  (\ref{position_continuity}) and (\ref{orient_continuity}) are identically satisfied and the rod model reduces to two static equilibrium equations, namely
\begin{eqnarray}
\frac{d\vec{f}}{ds} + \vec{\kappa} \times \vec{f} &=& - \vec{F}, \label{static_linear}\\
\frac{d\vec{q}}{ds} + \vec{\kappa} \times \vec{q} &=& \vec{f} \times \vec{r}- \vec{Q}
\label{static_angular}
\end{eqnarray}
in three vector unknowns, namely internal force $\vec{f}(s)$, internal moment $\vec{q}(s)$ and curvature vector $\vec{\kappa}(s)$. Since we assume that the rod is inextensible and unshearable, $\vec{r} = \hat{t}$, a known unit vector, which remains unchanged from reference (undeformed) configuration to deformed configuration as seen from the cross-section-fixed reference frame $\hat{a}_{i}(s,t)$. We also recall that in equations (\ref{static_linear}) and (\ref{static_angular}), the derivatives are all relative to the cross-section-fixed reference frame $\hat{a}_{i}(s,t)$.

\subsection{State-Space Form of Static Rod Model}
\label{state space form}
To use the above static rod model equations, (\ref{static_linear}) and (\ref{static_angular}), in two-step estimation method, we refer them  to cross-section-fixed reference frame $\hat{a}_{i}(s)$ and write them in state-space form
\begin{eqnarray}
\frac{dx}{ds} = f(x,u) + w, \hspace{1 pc} s \in [0,L], \label{statespace}
\end{eqnarray}
by defining the state $x(s) \in \mathbb{R}^{6}$ as components of $\vec{f}(s)$ and $\vec{q}(s)$ along $\hat{a}_{i}(s)$, input $u(s)  \in \mathbb{R}^{3}$ as components of $\vec{\kappa}(s)$ along $\hat{a}_{i}(s)$ and disturbance $w(s) \in \mathbb{R}^{6}$ as components of $\vec{F}(s)$ and $\vec{Q}(s)$ along $\hat{a}_{i}(s)$. We recognize that a causal relationship need not exist between input $u(s)$ and state $x(s)$. In fact, the constitutive law (\ref{const}) imposes an algebraic constraint between input $u(s)$ and state $x(s)$
\begin{equation}
\Phi \left( x, u, s \right) = 0, \hspace{1 pc} \Phi  \in \mathbb{R}^{6} \times \mathbb{R}^{3} \times \mathbb{R} \rightarrow\mathbb{R}^{3},
\label{phi}
\end{equation}
which for a repeating atomistic pattern will have a periodicity of $\lambda$ equal to the repeat interval and therefore $\Phi \left( x, u, s \right) = \Phi \left( x, u, s+\lambda \right)$. Note that the state-space equation (\ref{statespace}) is nonlinear, and moreover the state and input must satisfy the nonlinear algebraic constraint (\ref{phi}). We note that the state vector $x(s)$ represents the force and the moment, the input $u(s)$ represents the curvature vector and the disturbance $w(s)$ represents distributed load on the cantilever. We also note that the state vector $x(s)$ is prescribed at the free end of the cantilever (at $s=0$), thus $x(0)$ serves as initial conditions for the state-space equation (\ref{statespace}) .

If the constitutive law (\ref{phi}) were known, the above state-space equations could be solved as a constrained initial value problem using a Differential Algebraic Equation (DAE) solver. Once the unknown curvature vector $\vec{\kappa}(s)$ is solved, it can be further integrated to calculate the deformed shape of the rod respecting the position and orientation of the clamped end (at $s=L$) as summarized in \ref{curvature2shape}. But since the constitutive law is unknown, we outline the following two-step method for estimating the constitutive law. We assume that the only measurements available from the discrete structure simulations are the reference (undeformed) and deformed configurations of the cantilever, distributed loading $w(s)$ and the loading $x(0)$ at the free end. Several other estimation scenarios based on measurement contingencies are discussed in \citep[Table 1]{palanth:10a}.

\subsection{Two-Step Estimation Method}
\label{two step}

\begin{figure}[h]
\centering
\includegraphics[scale=0.55]{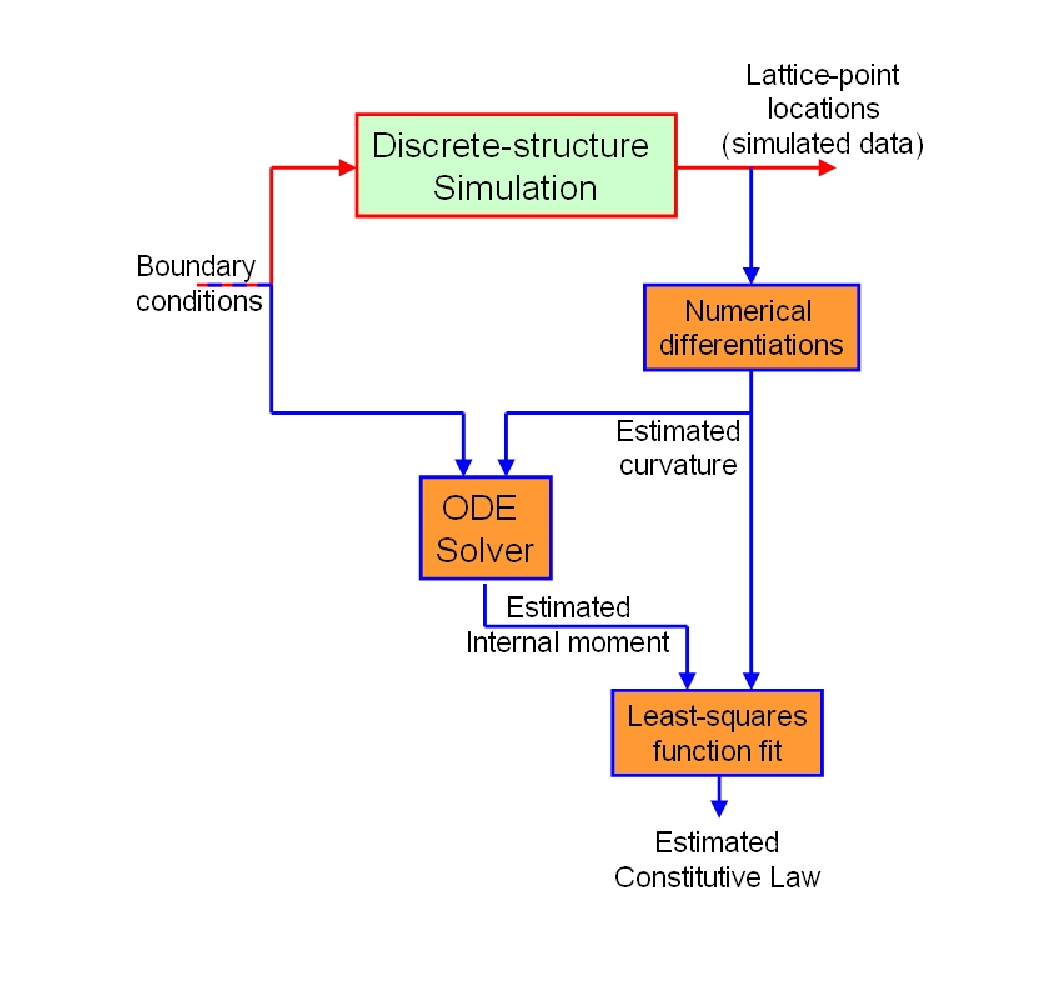}
\caption{Flowchart illustrating the components of the estimation of the constitutive law. The red arrows represent the path of information flow for a forward model simulation, while the blue arrows represent the path of information flow for inverse modeling to estimate the constitutive law. As seen from the chart, the inverse modeling framework uses data (or simulated data) and model information to estimate the constitutive law. For the results presented in this paper, the blocks shaded in green are implemented in Hyperworks Motionview, while the blocks shaded in brown are implemented in MATLAB Simulink.\label{flow}}
\end{figure}

In order to estimate the constitutive law (\ref{phi}), we first need to estimate its arguments, the state $x(s)$ and the input $u(s)$ from available measurements of a deformed state. Then, in the second step, we use function-approximation tools such as tools based on least-square fitting to identify the functional relationship (\ref{phi}). The first step, in general involves state-estimation and input-reconstruction methods depending on the available measurements \citep[Table 1]{palanth:10a}. In this paper, we have position data as an available measurement. So, we derive the input $u(s)$ (or equivalently, the curvature vector $\vec{\kappa}(s)$) from the position data using equation (\ref{kappa_n_R}) as explained next in Section \ref{shape2curvature}) and then solve the state-space equation (\ref{statespace}) for the state $x(s)$. Thus in this scenario, we do not use state-estimation or input-reconstruction methods. Figure~\ref{flow} illustrates the implementation of two-step estimation method in our scenario. 

We note that with the input $u(s)$ known, equation (\ref{statespace}) becomes an unconstrained initial value problem. We assume that accurate measurements of the initial condition $x(0)$ (loading at the free end) and disturbance $w(s)$ (distributed loading) are available. Difficulties due to inaccurate/ incomplete measurements can be circumvented by leveraging partial measurements of state $x(s)$ with unscented Kalman filter and/ or unscented unbiased minimum-variance (UUMV) filter as described in \citep{palanth:10a}.

\subsection{Estimation of Curvature Vector from Position Data}
\label{shape2curvature}

We use the position data of the reference (undeformed) and deformed configurations to estimate the input $u(s) = \left\{\vec{\kappa}(s)\right\}$. Here we introduced the notation $\left\{\vec{\kappa}(s)\right\}$ to denote the $3 \times 1$ column matrix representing the components of the physical vector $\vec{\kappa}(s)$ along $\hat{a}_{i}(s)$:

\begin{eqnarray}
\left\{\vec{\kappa}\right\} \isdef \left\{\begin{array}{c} {\kappa}_{1}\\ {\kappa}_{2}\\ {\kappa}_{3} \end{array}\right\} \isdef \left\{\begin{array}{c} \vec{\kappa}\cdot\hat{a}_{1}\\ \vec{\kappa}\cdot\hat{a}_{2}\\ \vec{\kappa}\cdot\hat{a}_{3} \end{array}\right\}. \label{kappa_comp}
\end{eqnarray}

To derive $\left\{\vec{\kappa}(s)\right\}$ from the position data, we first recall from \citep[Appendix 2]{goyal:05b} that

\begin{eqnarray}
\frac{d\left[L\right]}{ds} &=& -\left[\tilde{\kappa}\right]\left[L\right], \label{kappa_n_L}
\end{eqnarray}
where, the $3 \times 3$ square matrix

\begin{eqnarray}
 \left[L(s)\right] \isdef \left[\begin{array}{ccc} \hat{e}_{1}\cdot\hat{a}_{1} & \hat{e}_{2}\cdot\hat{a}_{1} & \hat{e}_{3}\cdot\hat{a}_{1} \\
\hat{e}_{1}\cdot\hat{a}_{2} & \hat{e}_{2}\cdot\hat{a}_{2} & \hat{e}_{3}\cdot\hat{a}_{2} \\
\hat{e}_{1}\cdot\hat{a}_{3} & \hat{e}_{2}\cdot\hat{a}_{3} & \hat{e}_{3}\cdot\hat{a}_{3} \end{array}\right] \label{transformation}
\end{eqnarray}
represents the transformation from the inertial reference frame $\hat{e}_{j} (s)$ (where $j = 1,2,3$) to the cross-section-fixed reference frame $\hat{a}_{i}(s)$ and

\begin{eqnarray}
\left[\tilde{\kappa}\right] \isdef \left[\begin{array}{ccc} 0 & -{\kappa}_{3} & {\kappa}_{2} \\
{\kappa}_{3} & 0 & -{\kappa}_{1} \\
-{\kappa}_{2} & {\kappa}_{1} & 0 \end{array}\right] \label{kappa_tilde}
\end{eqnarray}
is 3$\times$3 skew-symmetric matrix such that if vector $\vec{c} = \vec{\kappa}\times \vec{b}$ is the cross product of $\vec{\kappa}$ with any vector $\vec{b}$, then $\left\{\vec{c}\right\} = \left[\tilde{\kappa}\right]\left\{\vec{b}\right\}$. The result (\ref{kappa_n_L}) is derived in \citep[Appendix 2]{goyal:05b} from the Serret-Frenet formula
\begin{equation}
\left(\frac{d\hat{a}_{i}}{ds} \right)_{\hat{e}_{j}}= \vec{\kappa}  \times \hat{a}_{i} \ , \ \mbox{where recall that} \ i = 1,2,3 \label{serret-frenet}
\end{equation}
and where the subscript ${\hat{e}_{j}}$ denotes that the derivative is relative to the inertial reference frame ${\hat{e}_{j}}$.

Next, let the square matrix $\left[R(s)\right]$ represent the rotation transformation\footnote{Any change in the orientation of a cross-section from the undeformed configuration to the deformed configuration can be accomplished by a single Euler rotation about a unit vector $\hat{u}$ by angle $\theta$. In terms of the single Euler rotation parameters, $\left[R\right] = \exp{\left(\theta\left[\tilde{u}\right]\right)} = \left[I\right]+\left[\tilde{u}\right]\sin{\theta}+\left[\tilde{u}\right]^2\left(1-\cos{\theta}\right)$, where $\left[I\right]$ is the identity matrix and $\left[\tilde{u}\right]$ is the skew-symmetric matrix corresponding to $\hat{u}$.} of cross-section at $s$ (or equivalently, of the cross-section-fixed frame $\hat{a}_{i}(s)$) from reference configuration to deformed configuration. Then $\left[R\right]=\left[L_{ref}\right] \left[L_{def}\right]^{T}$ and we derive $\left[\tilde{\kappa}(s)\right]$  from equation (\ref{kappa_n_L}) in terms of $\left[R(s)\right]$ as

\begin{eqnarray}
\left[\tilde{\kappa}\right]  = \left[R\right]^{T}\left(\frac{d\left[R\right]}{ds} + \left[\tilde{\kappa}_{0}\right]\left[R\right]\right). \label{kappa_n_R}
\end{eqnarray}
The superscript $T$ denotes matrix transpose, subscripts $ref$ and $def$ correspond to reference and deformed configurations respectively, and $\left[\tilde{\kappa}_{0}(s)\right]$ is skew-symmetric matrix corresponding to the stress-free (reference) curvature  $\vec{\kappa}_{0}(s)$. Note that the stress-free curvature  $\vec{\kappa}_{0}(s)$ depends on the choice of the cross-section-fixed reference frame $\hat{a}_{i}(s)$ and satisfies equations (\ref{kappa_n_L}) and (\ref{serret-frenet}) in reference configuration (with $\left[L\right] =\left[L_{ref}\right]$).

Finally, we recognize that if we estimate the rotation $\left[R(s)\right]$ from the reference configuration to the deformed configuration, equation (\ref{kappa_n_R}) provides the curvature vector $\vec{\kappa}(s)$ by calculating the skew-symmetric matrix $\left[\tilde{\kappa}(s)\right]$, which has the three components of the curvature vector $\vec{\kappa}(s)$ along $\hat{a}_{i}(s)$ as per the equations (\ref{kappa_comp}) and (\ref{kappa_tilde}). There are several conceivable ways of estimating the rotation $\left[R(s)\right]$ from the reference and the deformed configuration position-data. One approach is to select a point cloud (e.g. cluster of atoms) from the discrete structure around each cross-section and estimate the best fit of $\left[R(s)\right]$ for the point cloud from the reference configuration to the deformed configuration. There are several commercially available tools (e.g. NX Imageware (UGS, Plano, TX)) to estimate the rigid body rotations of point clouds.

An alternative approach to derive the curvature vector $\vec{\kappa}(s)$ from position data is to use the Serret-Frenet formula (\ref{serret-frenet}) directly. This approach involves book-keeping of the cross-section-fixed reference frame $\hat{a}_{i}(s)$ in deformed configuration from position data. We note that either approach involves numerical differentiations, which may potentially be ill-posed and amplify noise of the measured data \citep{ramm:04a}. The effects of noise in the two-step estimation method with appropriate unscented filters is analyzed in \citep{palanth:10a}.
 
\section{Illustrative Results}
\label{results}

In this section, we illustrate the applicability of the two-step estimation method on a filament with an artificial discrete-structure with planar bending. We first describe the artificial discrete-structure in Sub-section \ref{discrete structure}. Next, in Sub-section \ref{simplifications}, we exploit some simplicities in the expected form of the constitutive law that are physically obvious from the discrete structure. In Sub-section \ref{planar rod}, we describe the state-space form of the reduced-order static rod model in the plane of bending. Finally, we present and discuss the estimation and validation results in Sub-sections \ref{estimation results} and \ref{validation results} respectively.

\subsection{An artificial discrete-structure}

\label{discrete structure}
\begin{figure}[h!]
\centering
\includegraphics{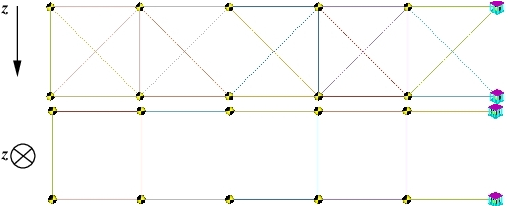}
\caption{Discrete structure of the filament. Two orthogonal views are shown, one along $z$-axis (top) and the other into the $z$-axis (bottom). Unit point masses are arranged in a cubic arrangement and five repeats of the cubic arrangement are shown. Each edge of the cubic cell has unit length and has linear springs of unit stiffness connecting the point masses in the undeformed (reference) configuration. In addition, the diagonal springs are placed on two opposite faces of the cube shown in the view along $z$-axis (top), but not on the other four faces of the cube.}
\label{unitlattice}
\end{figure}

We construct and simulate the discrete structure in a commercial, multi-body-dynamics software, Hyperworks MotionView. We follow the default system of units in Hyperworks, which is [Newton, Millimeter, Kilogram, Second]. In the discrete structure, we have thirty repeats of a cubic arrangement of unit point masses connected by linear springs of unit stiffness. Figure~\ref{unitlattice} shows two orthogonal views of a representative portion (five repeats) of the discrete structure in undeformed (reference) state. The springs are along each edge of the cube as well as along diagonals of two opposite faces as shown. Each edge of the cube is of unit length, so, the length of the entire filament $L = 30$ mm.

\subsection{Simplified constitutive law with in-plane bending}
\label{simplifications}

Before estimating the effective constitutive law, we first exploit the simplicities in its form that are physically obvious from the discrete structure. Since the discrete structure is uniform, we expect the effective constitutive law also to be uniform (homogeneous) along the length. In other words, there is no explicit dependence on $s$ in equation (\ref{const}). In addition, the constitutive-law equation (\ref{const}) also has no explicit dependence on the internal force $\vec{f}(s)$ for the chosen discrete structure \footnote{Although this simplicity may not be readily obvious, all the simplicities that we impose in this section will become plausible anyway with validation of the estimated constitutive law in Section \ref{validation results}}. So, the equation (\ref{const}) that needs to be estimated, simplifies to the following form:

\begin{equation}
\vec{\psi} \left( \vec{\kappa}, \vec{q} \right) = 0.
\label{const_simpler}
\end{equation}

Finally, we also note that due to the symmetries of the discrete-structure, the constitutive law is decoupled in the principal directions of bending and torsion. Choosing the cross-section-fixed reference frame $\hat{a}_{i}(s)$ along the principal directions, the vector equation (\ref{const_simpler}) simplifies to three scalar equations
\begin{equation}
\psi_{i} \left( \kappa_{i}, q_{i} \right) = 0,
\label{psi}
\end{equation}
where the subscript $i = 1, 2, $ or $3$ corresponds to each of the principal axes along $\hat{a}_{i}(s)$. Specifically, we let  $\hat{a}_{1}(s)$,  $\hat{a}_{2}(s)$, and  $\hat{a}_{3}(s)$ correspond to the first bending, second bending, and torsion axes, respectively, as illustrated along with a result in Figure \ref{case1}. The unit vector $\hat{a}_{3}(s)$ points along the outward normal to the cross-section. The unit vector $\hat{a}_{1}(s)$ is parallel to the faces of the cube containing diagonal springs, while $\hat{a}_{2}(s)$ is parallel to the faces not containing diagonal springs \footnote {Note that with this choice of reference frame, the stress-free curvature vector $\vec{\kappa}_{0}(s) = 0$ since the discrete structure is straight in undeformed (reference) state. This fact can be verified by satisfying Serret-Frenet formula (\ref{serret-frenet}) in the reference state.}. To present the viability of the two-step estimation method, we illustrate it for planar bending about only one principal bending axis, which we choose to be along $\hat{a}_{2}(s)$-axis.

\subsection{In-plane static rod model in state-space form}
\label{planar rod}

In-plane bending about the $\hat{a}_{2}$-axis occurs when a shear force is applied in the $\hat{a}_{1}$-direction and/ or bending moment is applied about the $\hat{a}_{2}$-axis. Referring all the physical vectors to the cross-section-fixed reference frame $\hat{a}_{i}(s)$, we note that the first and third components of $\vec{\kappa}(s)$ and $\vec{q}(s)$ and the second component of $\vec{f}(s)$ vanish for the planar bending. Furthermore, we will have no distributed load. In this case the governing vector equations (\ref{static_linear}) and (\ref{static_angular}) of the static rod model further reduce to the following three scalar nonlinear differential equations
\begin{eqnarray}
\frac{dq_{2}}{ds} &=& -f_{1} , \label{gov1}\\
\frac{df_{1}}{ds} &=& -f_{3} \kappa_{2}, \label{gov2}\\
\frac{df_{3}}{ds} &=& f_{1} \kappa_{2},  \label{gov3}
\end{eqnarray}
with the constitutive law to be estimated
\begin{eqnarray}
\psi_{2} \left( \kappa_{2}, q_{2} \right) =0. \label{law}
\end{eqnarray}
Equations (\ref{gov1}-\ref{gov3}) are in the state-space form (\ref{statespace}), as described in \citep{law}, with

\begin{eqnarray}
x = \left\{\begin{array}{c} {f}_{1}\\ {f}_{3}\\ {q}_{2}\end{array}\right\}, \ \ w = \left\{\begin{array}{c} 0\\ 0\\ 0\end{array}\right\} \ \ \mbox{and} \ \ u = {\kappa}_{2}.\label{statespace_reduced}
\end{eqnarray}

\subsection{Estimation of the constitutive law}
\label{estimation results}

\begin{figure}[h!]
\centering
\includegraphics[width=90mm]{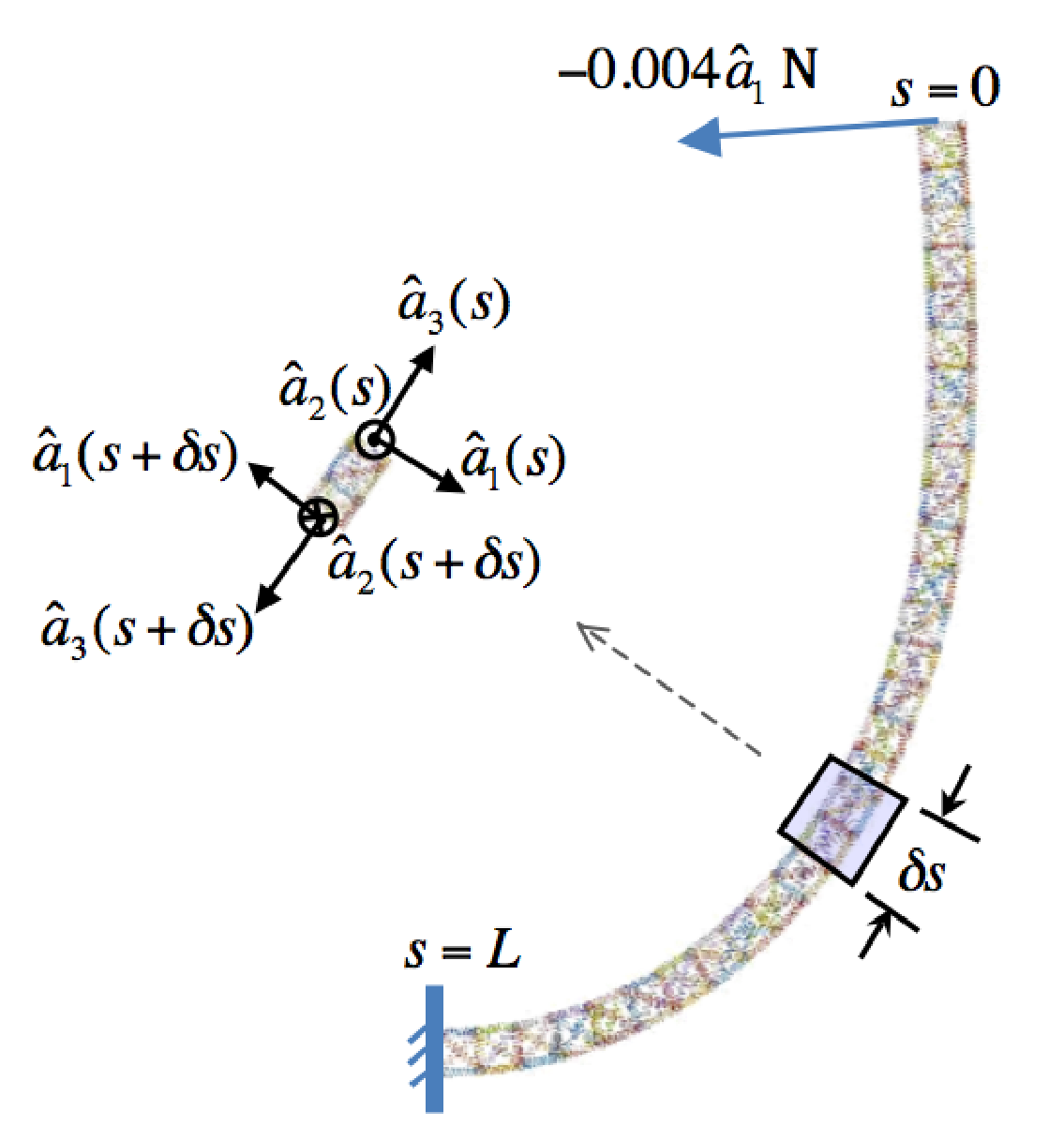}
\caption{Deformed shape of the discrete-structure cantilever in static equilibrium. The free end has a prescribed shear force, but has no tension and no bending moment. Since there are thirty cubic cells of 1 mm edge each, $L = 30$ mm. The cross-section-fixid reference frame $\hat{a}_{i}(s)$ is also shown. The unit vector $\hat{a}_{3}(s)$ points along the outward normal to the cross-section. The unit vector $\hat{a}_{1}(s)$ is chosen in the plane of bending, while $\hat{a}_{2}(s)$ is parallel to the axis of bending.}
\label{case1}
\end{figure}

\begin{figure}[h!]
\centering
\hspace{-0.1in}
\includegraphics[width=90mm]{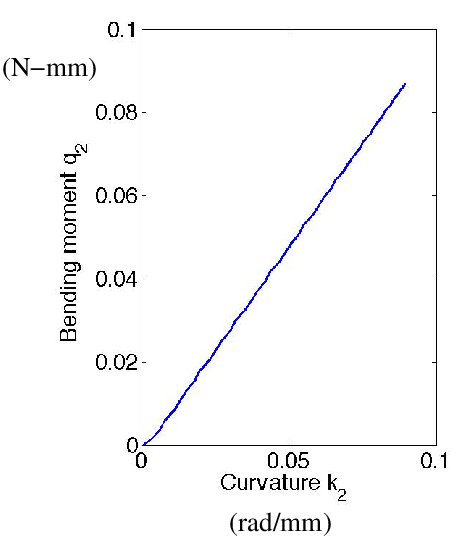}
\caption{Estimated constitutive law.}
\label{ConstitutiveLaw}
\end{figure}

For estimation of the effective constitutive law, we deform the cantilevered discrete-structure in Hyperworks MotionSolve by exerting a shear force $f_{1}(0)=-0.004$ N at the free end (at $s=0$) as shown in Figure~\ref{case1}. We simulate the static equilibrium configuration of the discrete structure using the Force Imbalance Method-Type D for the static analysis in Hyperworks MotionSolve. From the deformed configuration, we estimate the curvature $\kappa_{2}(s)$ as outlined in Section \ref{shape2curvature}. Next, to estimate the internal moment $q_2(s)$, we use the estimated curvature $\kappa_2(s)$ as a known input $u(s)$ in the state-space equations (\ref{gov1}) - (\ref{gov3}) of the in-plane static rod model. We solve the state-space equations (\ref{gov1}) - (\ref{gov3}) for $x(s)$, which contains $q_2(s)$, in MATLAB Simulink with the initial conditions:

\begin{eqnarray}
x(0) = \left\{\begin{array}{c} {f}_{1}(0)\\ {f}_{3}(0)\\ {q}_{2}(0)\end{array}\right\} \isdef \left\{\begin{array}{c} -0.004 \ \mbox{N}\\ 0 \ \mbox{N}\\ 0 \ \mbox{N-mm}\end{array}\right\}.\label{initial conditions}
\end{eqnarray}

Finally, to estimate the functional relationship (\ref{law}) between $\kappa_{2}$ and $q_{2}$ we express the function, $\psi_{2}$, as a sinusoidal-basis-function expansion, where the unknown coefficients of this expansion are then found by standard least-squares-fitting. Figure~\ref{ConstitutiveLaw} shows the estimated constitutive relationship between the bending moment and curvature.

\subsection{Validation of the constitutive law}
\label{validation results}

\begin{figure}[h]
\centering
\includegraphics[width=100mm]{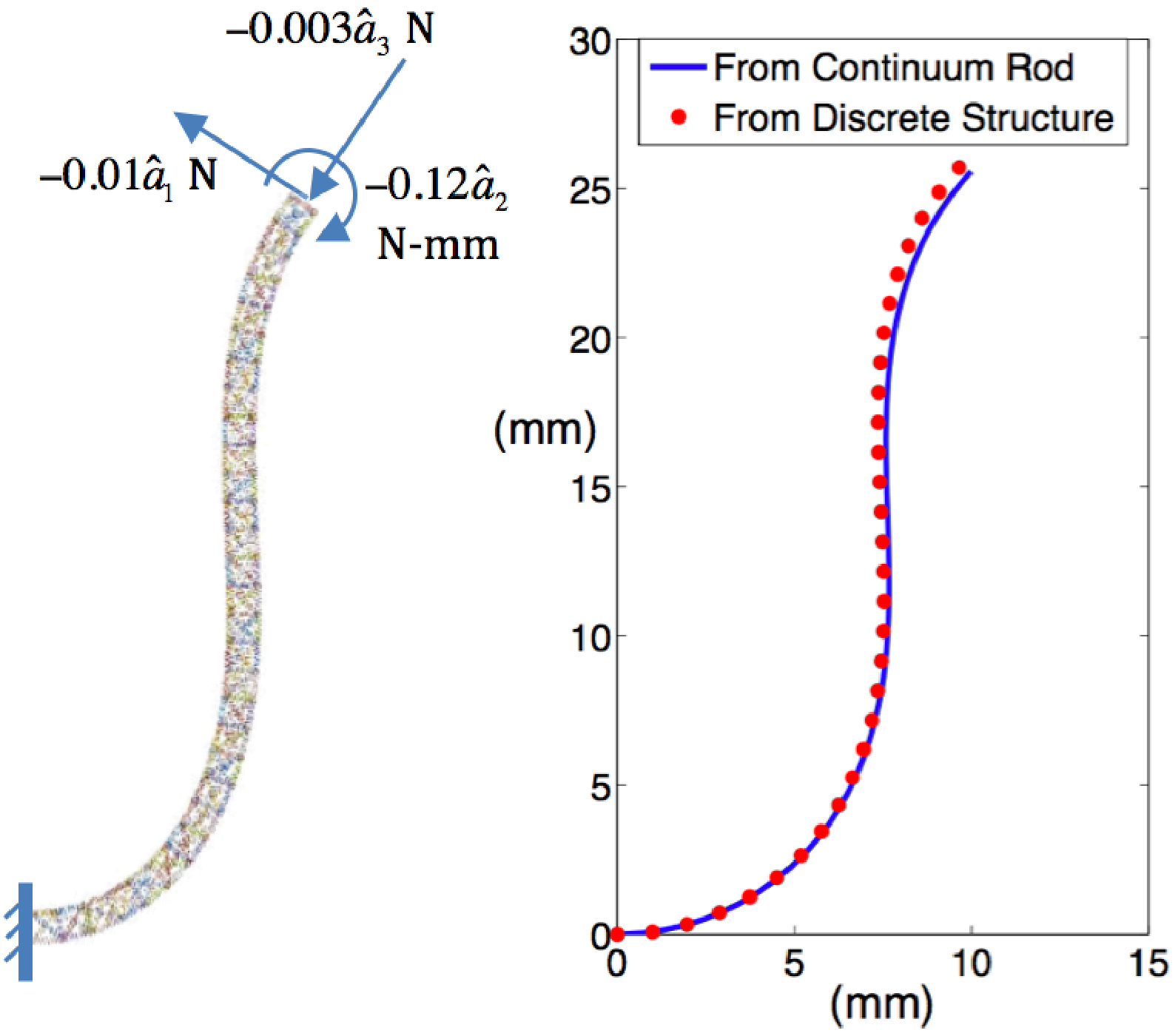}
\caption{Deformed shape of the discrete-structure cantilever in static equilibrium (left). The free end has prescribed shear force, tension as well as bending moment. Centerline predicted from discrete-structure simulation in Hyperworks (dots) and continuum-rod simulation (solid curve) employing the estimated constitutive-law (right).}
\label{case2}
\end{figure}

To validate the estimated constitutive-law, one may test it in several different loading environments (static or dynamic) and evaluate if it predicts the same deformation behavior from the rod model as does the discrete-structure simulation. Here we present one spot test in static equilibrium. The loading conditions for the spot test are shown in Figure~\ref{case2} on the deformed equilibrium of the discrete structure (on the left). In this case we apply shear force, compressive force as well as bending moment at the free end. We apply these loading conditions in the Simulink state-space-model of the continuum rod employing the estimated constitutive-law shown in Figure~\ref{ConstitutiveLaw}. We set it up as DAE in Simulink with constitutive law as the known, algebraic constraint. Figure~\ref{case2} shows (on the right) the comparison of the centerline predicted from the discrete-structure simulation in Hyperworks (dots) and the continuum-rod simulation (solid curve) employing the estimated constitutive-law. The closely matching shape validates the estimated constitutive-law.

\subsection{Analysis and discussion of results}
\label{analysis}

The close agreement obtained in Section \ref{validation results} of the centerline deformations of the discrete-lattice structure and the continuum rod-model in the spot-test loading not only illustrates the viability of the two-step estimation-method, but also corroborates several assumptions that we made along the way. In particular, we assumed inextensibility and unshearability in the rod model equations (\ref{gov1}) - (\ref{gov3}), while the discrete-structure simulations are inherently free from such assumptions. The slenderness ratio (length/thickness) of the discrete structure is 30. Therefore the results corroborate the assumptions of inextensibility and unshearability in Kirchhoff rod theory for slender filaments. Furthermore, the close agreement corroborates the simplifications that we imposed {\it a priori} on the constitutive law in Section \ref{simplifications}. The agreement is also noteworthy when considering that each simulation is run independently with different softwares and solvers, and therefore serve as a validation study for the softwares and solvers as well. The accuracy of the results obtained using commercial softwares and standard numerical methods is promising for the ease of implementing the method to a variety of applications involving filament-type deformable structures.

\begin{figure}[h]
\centering
\includegraphics[width=85mm]{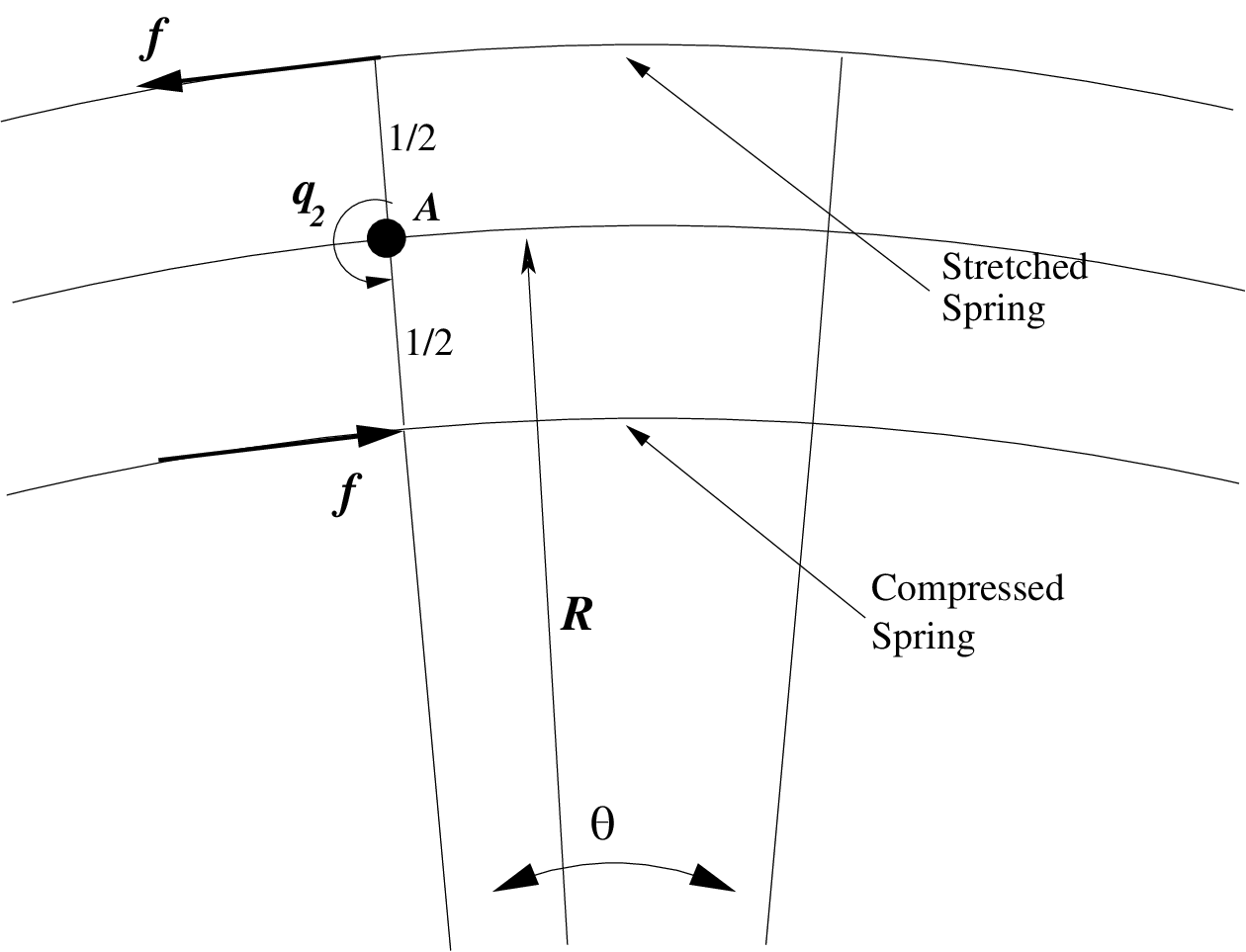}
\caption{A schematic to derive the constitutive law $\psi_{2} \left( \kappa_{2}, q_{2} \right) =0$ from the restoring spring-forces developed in the discrete structure. The neutral surface is approximately half way between the top and bottom surfaces.}
\label{explain}
\end{figure}

We recall here that the first-principle derivation of the constitutive law from atomistic-level interactions is often impractical in general. However, for the discrete structure that we chose, we indeed can corroborate the estimated constitutive law from an approximate first-principle derivation.  Consider the spring forces in the deformed discrete structure in Figure \ref{explain} showing one cubic cell. If $R$ is the radius of curvature of the neutral surface and $\theta$ is the bend angle, $R\theta = 1$ mm and $\kappa_2 = 1/R = \theta/$mm. The restoring moment $q_2$ is related to the restoring forces developed in the springs on the top and bottom faces of the cube. The top two springs in the plane of bending are stretched, while the bottom two springs are compressed. The  restoring force $f$ developed in each of the four springs is $\theta/2$ N and therefore, the resultant moment $q_2 = \theta$ N-mm. Eliminating $\theta$ between $q_2$ and $\kappa_2$, we get $q_2 =$ (1 N-mm$^2$) $\kappa_2$, which corroborates the nearly linear relationship between $q_2$ and $\kappa_2$ observed in Figure \ref{ConstitutiveLaw}.

\section{Conclusions}
\label{conclusions}

The continuum-rod model has emerged as an efficient tool to describe the long length and time scale dynamics of nonlinear deformations of microfilaments and sem-flexible biofilaments such as DNA. However, the usefulness of the rod model is severely limited by the lack of knowledge of an accurate constitutive law. Discrete-structure simulations, such as MD simulations of biomolecules such as DNA for small length and time scales can provide ample data for the accurate estimation of an effective constitutive law.

In this paper, we described a method based on a recently proposed two-step inverse approach \citep{palanth:10a} to estimate the constitutive law using discrete-structure simulations. We illustrated the applicability of the method on a simple representative example. In particular, we considered a cantilever filament with an artificial discrete-structure. We simulated its deformation in response to a prescribed loading using a multi-body dynamics (MBD) solver. We then estimated an effective constitutive law from the deformation data using the two-step method. Finally, we illustrated how the estimated constitutive law can be validated by employing it in the continuum-rod model and comparing the simulation results with those of discrete-structure simulations under an independent set of cantilever loading conditions. The method presented in this paper can leverage MD simulations of the discrete atomistic structure of microfilaments to estimate their constitutive laws.

\section{Acknowledgments}
\label{acknow}

The authors gratefully acknowledge Altair Eng., Inc. for providing their CAE tool Hyperworks for this research. The authors also gratefully acknowledge the training provided by Mr. Sundar Nadimpalli, Application Specialist - Multibody Dynamics, Altair Eng., Inc., Bangalore to construct and simulate a discrete structure model in Hyperworks MotionView. The authors are also grateful to Bettis lab for partially sponsoring the first author's travel to ASME IDETC 2009 conference in San Diego to present this work. The authors also acknowledge ASME Publications for permission to reproduce slightly modified versions of the results presented in the above-said conference \citep{hinkle:09a}. 


\appendix

\section{Shape from Curvature}
\label{curvature2shape}

\subsection{Three-dimensional deformation}
\label{3Dcase}


Here, we summarize how to solve for the centerline shape $\vec{R}(s)$ from the curvature components $\left\{\vec{\kappa}(s)\right\}$ for an inextensible and unshearable rod. Recall from Section \ref{rod model} that
\begin{eqnarray}
\left(\frac{d\vec{R}}{ds}  \right)_{\hat{e}_{j}} &\isdef& \vec{r}, \label{position gradient}
\end{eqnarray}
where the subscript ${\hat{e}_{j}}$ denotes that the derivative is relative to the inertial reference frame ${\hat{e}_{j}}$. For an inextensible and unshearable rod, the position gradient $\vec{r}(s) = \hat{t}(s)$. So, we can integrate equation (\ref{position gradient}) to get $\vec{R}(s)$ if we find

\begin{eqnarray}
\left\{\hat{t}\right\}_{\hat{e}_{j}} \isdef \left\{\begin{array}{c} \hat{t}\cdot\hat{e}_{1}\\ \hat{t}\cdot\hat{e}_{2}\\ \hat{t}\cdot\hat{e}_{3} \end{array}\right\}, \label{t_inert_comp}
\end{eqnarray}
the components of the unit tangent vector $\hat{t}(s)$ along ${\hat{e}_{j}}$. But

\begin{eqnarray}
\left\{\hat{t}\right\}_{\hat{e}_{j}} =  \left[L\right]^T\left\{\hat{t}\right\}, \label{t_transformation}
\end{eqnarray}
where $\left[L(s)\right]$ is the transformation matrix defined in equation (\ref{transformation}), the superscript $T$ denotes the transpose and the $3 \times 1$ matrix

\begin{eqnarray}
\left\{\hat{t}\right\} \isdef \left\{\begin{array}{c} \hat{t}\cdot\hat{a}_{1}\\ \hat{t}\cdot\hat{a}_{2}\\ \hat{t}\cdot\hat{a}_{3} \end{array}\right\} \label{t_body_comp}
\end{eqnarray}
represents the components of $\hat{t}(s)$ along ${\hat{a}_{i}}$. We know $\left\{\hat{t}\right\}$ a priori from the choice of ${\hat{a}_{i}}$. The transformation matrix $\left[L(s)\right]$ can be solved from $\left\{\vec{\kappa}(s)\right\}$ by integrating equation (\ref{kappa_n_L}). We use the following approximation to integrate equation (\ref{kappa_n_L}):

\begin{eqnarray}
\left[L(s+{\delta}s)\right] \approx \exp{\left(-\left[\tilde{\theta}\right]\right)}\left[L(s)\right], \label{L_update}
\end{eqnarray}
where $\left[\tilde{\theta}\right]$ is the skew-symmetric matrix corresponding to

\begin{eqnarray}
\left\{\vec{\theta}\right\} \isdef \displaystyle\int^{s+{\delta}s}_{s} \left\{\vec{\kappa}\right\}\,ds \label{incremental_rotation}
\end{eqnarray}
such that if vector $\vec{c} = \vec{\theta}\times \vec{b}$ is the cross product of $\vec{\theta}$ with any vector $\vec{b}$, then $\left\{\vec{c}\right\} = \left[\tilde{\theta}\right]\left\{\vec{b}\right\}$.

We evaluate the integral in equation (\ref{incremental_rotation}) using trapezoidal rule, integrate equation (\ref{kappa_n_L}) to solve for $\left[L(s)\right]$ using numerical approximation (\ref{L_update}), solve for $\left\{\hat{t}(s)\right\}_{\hat{e}_{j}}$ from equation (\ref{t_transformation}) and finally recognizing $\vec{r}(s) = \hat{t}(s)$, integrate equation (\ref{position gradient}) using trapezoidal rule to get the centerline curve $\vec{R}(s)$ of the deformed rod.

\subsection{In-plane bending}
\label{2Dcase}

For in-plane bending presented in Section \ref{results}, the shape integration from the curvature $\kappa_2$ is simpler than the above algorithm for the general 3-D case. If the plane of bending is designated as $x-y$ plane and if $\phi(s)$ is the angle that centerline tangent makes with the $x$-axis, then

\begin{eqnarray}
\frac{d\phi}{ds} &=& \kappa_2.  \label{kappa_2}
\end{eqnarray}
Furthermore, if the $x$ and $y$ components of $\vec{R}(s)$ are denoted by $R_x(s)$ and $R_y(s)$ respectively, then

\begin{eqnarray}
\frac{dR_x}{ds} = \cos{\phi} \ \ \mbox{and} \ \ \frac{dR_y}{ds} = \sin{\phi} . \label{RxRy}
\end{eqnarray}
Thus, we can integrate equation (\ref{kappa_2}) using trapezoidal rule to get $\phi(s)$ and then integrate equation (\ref{RxRy}) to get $R_x(s)$ and $R_y(s)$ respecting the boundary conditions at the clamped end (at $s=L$) of the cantilever. In particular, we can integrate along the increasing $s$-direction starting from the free end ($s=0$) with $\phi(0)$, $R_x(0)$ and $R_y(0)$ (the constants of integration) as unknowns and find their values satisfying the clamped values of $\phi(L)$, $R_x(L)$ and $R_y(L)$.

\bibliographystyle{model1-num-names}
\bibliography{JMPS}







\end{document}